\documentclass[aps,prl,twocolumn,showpacs]{revtex4-1}
\usepackage{graphicx} 
\begin{document}
\title
{\bf A Measurable Force driven by an Excitonic Condensate in DQWs}
\author{T. Hakio\u{g}lu$^{\bf (1,2)}$, Ege \"{O}zg\"{u}n$^{\bf (1)}$, Mehmet G\"{u}nay$^{\bf (1)}$}
\affiliation{
${\bf (1)}$ {Department of Physics, Bilkent University, 06800 Ankara, Turkey}
\break
${\bf (2)}$ {Institute of Theoretical and Applied Physics, 48740 Turun\c{c}, Mu\u{g}la, Turkey}
}  
\begin{abstract}
New free energy related signatures of the condensed excitons in Double Quantum Wells (DQW) are predicted and experiments are proposed to measure the effects. These signatures are related to the measurement of a conceptually new kind of force ($\approx 10^{-9} N$) due to the condensate. This force, which may be coined as the Exciton Condensate (EC)-force is attractive and reminiscent of the Casimir force between two perfect metallic plates, but also distinctively different from it by its driving mechanism and dependence on the parameters of the condensate. The proposed experiments here are based on a recent experimental work on a driven micromechanical oscillator with a proven high quality factor. The free energy related measurements are immune to the commonly agreed drawbacks of the existing photoluminescence experiments. In this regard, the proposed experiments are highly decisive about the EC.
\end{abstract}
\maketitle
In the late 40's Casimir predicted the presence of an unusual attractive force between two uncharged metallic plates held in vacuum\cite{CF}. The Casimir force (CF) is a pure quantum mechanical effect manifested by the vacuum fluctuations of the electromagnetic field in the presence of metallic boundaries. The CF is attractive between the two ideally infinite metallic plates and the force per unit area (the Casimir pressure) is given by $P_c={\cal F}_c/A=-\pi^2 \hbar c/(240 D^4)$ where $A$ is the area of the plates and $D$ is the separation between them. For typical values $A\simeq 1\mu m^2$ and $D\simeq 100 \AA$, ${\cal F}_c \simeq -1.3 \times 10^{-7} N$. The early measurements of Casimir force were done between a metal plate and a metal sphere\cite{early_Casimir}. The earliest successful measurements, on the other hand, between two metallic plates were more demanding due to the requirement of the perfect alingment between the plates. These experiments date back to eleven years ago where a perfect agreement with the theory was achieved.\cite{BCOR}.      

There are many equivalent ways of deriving the Casimir force in arbitrary geometries\cite{Casimir_theoretical_ways}. One way is to calculate the difference between the vacuum free energies and their dependence on the geometry. Since the free energy as well as the photon pressure is infinite for the electromagnetic vacuum, one has to apply delicate regularization techniques to extract the finite part of the photon pressure difference between the sides of the metallic plates. On the other hand, this phenomenon can be generalized to a wider conceptual basis if one replaces the electromagnetic vacuum with the many body ground state of an interacting quantum system. The idea of the vacuum can then be extended to a more general context including for instance, the macroscopically coherent ground states. The role of the vacuum fluctuations are then played by the critical fluctuations of the order parameter describing the macroscopic order. In confined systems these critical fluctuations are geometry dependent and quantized through the boundary conditions. Therefore it is expected that some statistical mechanical systems at the boundary of a phase transition also exhibit a similar effect. Indeed, this is known as the critical Casimir Force (CCF) which was predicted\cite{FG} and measured\cite{HHGDB} for binary liquid mixtures. The CCF, was also theoretically predicted for a BEC\cite{BEC_CF_Th} but experimentally has not been verified yet. On the other hand, the Casimir-Polder-like force between a BEC and a semiconductor plane has been measured\cite{BEC_CF_Exp}. 

The free energy of an interacting many body ground state may also depend on the geometrical scale through various other factors than the boundary conditions discussed above. One distinct case is a condensate of which the pairing interaction strength depends on the physical size of the system (see below). An example is spatially indirect coupling between interacting electron-hole systems confined to two separate layers (the DQW geometry). If the layer separation is on the order of an exciton Bohr radius $a_B$, the electrons and holes in separate layers form bound pairs, i.e. Wannier Mott excitons. In sufficiently low densities, when excitons act like independent bosons, they are expected to Bose-Einstein condense, below a $T_c$, inversely proportional to their reduced mass, into a phase coherent ground state\cite{EC_general} with a gap opening in the energy spectrum. Due to the small exciton reduced mass, the critical temperature of the exciton condensate should be on the order of a few Kelvin. As the density increases, the exciton wavefunctions start spatially overlapping, with a higher Fermi energy scale than the pairing interactions, moving into a BCS like paired ground state. A thorough understanding of this mechanism has not been conclusive mostly due to the inherent difficulties\cite{EC_interactions_general} such as the short exciton lifetime, the momentum as well as angular momentum dependence of the residual interactions between the fermionic constituents\cite{Combescot} and unconventional pairing symmetries\cite{1}. Commonly used experimental techniques such as the photoluminescence measurements are also limited in probing all components of the condensate due to the angular momentum selection rules\cite{EC_interactions_general}. It should be realized that EC research with such theoretical and experimental challenges is a broad resource in better  understanding the unconventional aspects of many body interacting quantum systems in general. It is a common sense to say that, in such a broad area, a rich variety of experimental methods going beyond the photoluminescence techniques should avail. The free energy related measurements, which we discuss here, are among such alternatives. 

The Coulomb pairing interaction strength in an interacting electron-hole system in a DQW geometry is an exponential function of the layer separation $D$ between the electron and the hole layers given by $V_{eh}({\bf q})=-e^2/(2\varepsilon q)\,e^{-q\,D}$. Here $q=\vert {\bf q}\vert$ is the inplane exchange momentum between the electrons and the holes and $\epsilon$ is the dielectric constant of the medium. The s-like conduction band in one of the layers and the p-like valence band in the other layer of the DQW structure are populated by electrons and holes using a pump laser resonant with the semiconductor gap. Under lattice temperatures of a few Kelvin, a few nanoseconds is sufficient for the electrons and the holes to relax with the lattice. In experiments, the lifetime of the created excitons can be promoted up to a few $\mu s$ if a perpendicular electric field is applied to minimize the dipolar coupling. However, even a marginal dipole strength is sufficient to couple with the spontaneous radiation field. The bright excitons with their odd angular momenta couple to the radiation field and recombine within a few $\mu s$. It is commonly accepted that, due to this radiative dipolar coupling, the ground state of the condensed excitons is dominated  by the dark states with bright dominated states existing at a slight higher energy scale above the ground state, on the order of 10 $\mu eV$. Despite this statement has been widely accepted, it was adopted from our knowledge in the normal excitonic fluid, not the condensate. From the condensate's perspective, this statement was verified recently\cite{2} starting from a microscopic theory including the coupling of the bright state with the radiation field. On the other hand, there is a faster mechanism for the ground state to be populated by dark excitons, i.e. the Pauli exchange scattering of two bright states into two dark ones\cite{Combescot}. This happens when two bright excitons exchange their electrons or holes within their proper QWs. Unless the density is extremely low, this mechanism is only weakly dependent on the exciton density and should occur in a wide density range. Although the bright-to-dark and dark-to-bright transitions are fundamental microscopic processes of equal strengths, under low temperature conditions more bright states turn into the dark ones in order to lower the free energy, yielding a dominantly dark ground state. The basic reason behind the lack of complete experimental evidence of the ground state in EC is that dark states do not couple to light due to their even angular momenta. 

In this letter, we demonstrate that the free energy related experiments on the condensed excitons do not differentiate between the dark and the bright components and hence, offer a promising complementary method to the photoluminescence measurements that can only probe the bright contribution. In these systems, the free energy depends, through the condensate's order parameter, on the layer separation $D$. For smaller $D$, the coupling is stronger and the condensation free energy should be lower, pointing at a new type of an attractive force between the plates of the DQW, as a signature of the condensation. This force may be coined as the EC-force and we address in this letter three  fundamental questions: 1) Can we understand the analytic dependence of the EC-force on the physical parameters such as $D$ and the exciton concentration $n_x$?, 2) Is the EC-force  measurable under realistic conditions and current experimental accuracy? 3) If so, can new experimental methods be proposed for its measurement? 

In a recent work, the free energy of an EC was numerically calculated using a self-consistent mean field Hartree-Fock scheme\cite{2}. The calculations indicate the existence of a sharp phase boundary separating the condensed phase from the fluctuating normal liquid. In EC one has a large number of controllable parameters like the density of condensed excitons $n_x$, the electron-hole density imbalance $n_-$, the layer separation $D$ and the temperature $T$. The sharp phase boundary, separating the EC from the fluctuating normal liquid, is defined by the critical values of these parameters. It was predicted that\cite{2}, in EC the phase boundary is present even at zero temperatures, where the free energy has a sharp second order phase transition defined by the remaining set of critical parameters. At zero temperature, the numerical solution of the free energy resembles the shape of an inverted parabola as a function of $D_c-D$ near $D\simeq D_c$ where $D_c$ is the critical layer separation for fixed $n_x$ and $n_-$. Due to the attractive coupling and the negative condensation energy, the EC-force exists not only at, but also deep inside the phase boundary. In this work, we provide an order of magnitude estimation of this force with its analytical dependence on the critical parameters.    

In a many body condensate, the condensed phase is determined by the condensation energy given by $\Delta \Omega=\Omega_{\Delta}-\Omega_{N} \le 0$, where the equality holds for the phase boundary. Here $\Omega_{\Delta}$ and $\Omega_{N}$ are the total free energies in the condensed and the uncondensed phases respectively. In an EC, we observe two types of dependence on $D$. The first is the critical thermal fluctuations of the condensate around its mean value near $T_c$. This term is suppressed if $T \ll T_c$. The second dependence on $D$ arises from the condensate's dominantly dark order parameter $\Delta_{\bf k}$ where ${\bf k}=(k_x,k_y)$ is the planar quasiparticle momentum (see \cite{2,1}). The dependence of $\Delta_{\bf k}$ on the layer separation is essential for the predictions in this work. On the other hand, one also has to understand the role of the electron and hole self energies. These are driven by the $D$-independent intralayer Coulomb interaction. Our numerical calculations indicate that although the inclusion of the self energies and the realistically different electron and hole masses perturbatively shift the position of the phase boundary, the existence and the shape of it is unaffected. In this letter, we also ignore the thermal fluctutations of the order parameter by formulating the theory at zero temperature.

The EC-force arises from the dependence of the condensation energy $\Delta \Omega$ on $D$ through $\Delta_{\bf k}$ as given by
\begin{equation}
{\cal F}_{EC}=-\frac{\partial \Delta \Omega}{\partial \Delta_{\bf k}} \frac{\partial \Delta_{\bf k}}{\partial D}~.
\label{Casimir_force}
\end{equation}
with a sum over ${\bf k}$ implied. The free energy is given up to a constant term by the standart expression at finite temperatures,   
\begin{eqnarray}
\Omega_{\Delta}=\sum_{\sigma,{\bf k}}\,\Bigl\{\sigma f_\sigma \frac{\Delta_{\bf k}^2}{(2E_{\bf k})}-\frac{\partial}{\partial \beta}\,\ln(1-f_\sigma)\Bigr\} 
\label{FE_1} 
\end{eqnarray} 
Here $\sigma=(+,-)$ denotes the upper and the lower excitonic branches separated by condensate's gap and at zero temperature only the lower branch contributes. Additionally, $\beta=1/(k_B T)$, $\Delta_{\bf k}$ is the order parameter of the condensate\cite{1}, $f_{\sigma}=1/[1+e^{\beta(\epsilon_{\bf k}^{(-)}+\sigma E_{\bf k})}]$ is the Fermi-Dirac factor for the quasi-particles in the energy branch $\sigma$, $\epsilon_{\bf k}^{(-)}=(\xi^{(e)}_{\bf k}-\xi^{(h)}_{\bf k})/2$ and $\xi^{(e)}_{\bf k}=\hbar^2{\bf k}^2/(2m_e)-\mu_e$, $\xi^{(h)}_{\bf k}=\hbar^2{\bf k}^2/(2m_h)-\mu_h$ parameterized by the electron and the hole band masses $m_e, m_h$ and the chemical potentials $\mu_e, \mu_h$, and the quasiparticle eigenenergies $E_{\bf k}=(\epsilon_{\bf k}^2+\Delta^2_{\bf k})^{1/2}$ where $\epsilon_{\bf k}=(\xi^{(e)}_{\bf k}+\xi^{(h)}_{\bf k})/2$. Eq.(\ref{FE_1}) reduces, at zero temperature, to the standart free energy of the condensed state given by $\Omega_{\Delta}=-\sum_{{\bf k}}\,\Bigl\{\Delta_{\bf k}^2/(2E_{\bf k})+E_{\bf k}\Bigl\}$. In what follows we assume that the electron and the hole masses are equal\cite{expl_2}, whereas allow, for now, an imbalance between their concentrations. We have then, $\epsilon_{\bf k}^{(-)}=-\mu_-$ where $\mu_-=(\mu_e-\mu_h)/2$. 

At zero temperature, and in the light of the previous discussions, $\Delta_{\bf k}$ is given by\cite{2}     
\begin{eqnarray} 
\Delta_{\bf k}= -\frac{\pi e^2}{\varepsilon}\int\frac{d{\bf q}}{(2\pi)^2} \frac{e^{-q\,D}}{q}G_{{\bf k}+{\bf q}}\, 
\label{op_0}
\end{eqnarray}
where
\begin{eqnarray}
G_{\bf k}=\frac{\Delta_{\bf k}\,F(\epsilon_{\bf k}^{(-)},E_{\bf k})}{\sqrt{(\epsilon_{\bf k}-\mu_x)^2+\Delta_{\bf {k}}^2}}\,. 
\label{op_1}
\end{eqnarray}
Here $F(\epsilon_{k}^{(-)},E_{k})=\Theta(-\epsilon_{k}^{(-)}- E_{k})-\Theta(-\epsilon_{k}^{(-)}+ E_{k})$ with $\Theta(x)$ describing the step function. The solution of Eq.(\ref{op_0}) should be found numerically together with the constraints on $n_x$ and $n_-$ fixing the exciton chemical potential $\mu_x=(\mu_e+\mu_h)/2$ and imbalance chemical potential $\mu_-$.  

\begin{figure}
\includegraphics[scale=0.34,angle=0]{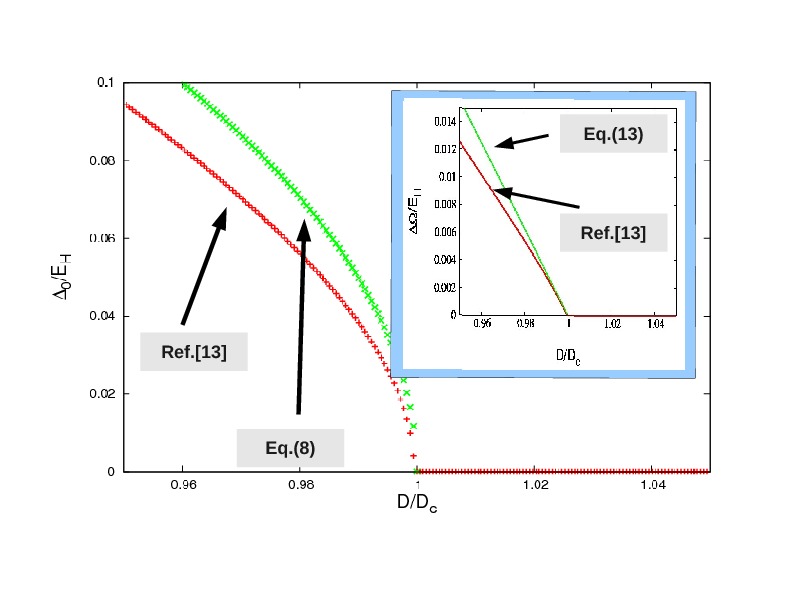}
\vskip-0.7cm
\caption{(Color online) Comparison of the numerical calculations in \cite{2} with Eq.(\ref{around_D_c}) and Eq.(\ref{Omega_at_D_c}) as $D$ is varied near $D_c$. The main figure depicts $\Delta_0$ and the inset is the condensation energy $\Delta \Omega$. Both energy scales are normalized with the Hartree energy $E_H=e^2/(4\pi \epsilon a_B)$ where $a_B=\hbar^2 4\pi \epsilon/(e^2 m_x)$ is the exciton Bohr radius. Note that the numerical solution includes electron and hole self energies as well as their realistically different masses.}
\label{check_Delta_0_Omega_0}
\end{figure}

In order to understand the EC-force given by Eq.(\ref{Casimir_force}) one has to understand the dependence of the condensate's order parameter analytically on the layer separation $D$. For such a model with a predictive power, the basic features of the earlier numerical results should also be reproducible. On the other hand, analytic calculations here are more demanding than those in the constant interaction models in atomic condensates by the presence of the momentum dependent interaction $V_{eh}({\bf q})$. Due to the lucky presence of the exponential in Eq.\,(\ref{op_0}), the leading contribution comes from the region $q \ll 1/D$. Expanding $G_{{\bf k}+{\bf q}}$ upto second order in $q$ near $q=0$, i.e. $G_{{\bf k}+{\bf q}} \simeq G_{\bf k}+\nabla_{\bf k} G_{\bf k}.{\bf q}+G_{\bf k}^{\prime\prime}q^2/2$ where $G_{\bf k}^{\prime\prime}$ is the second derivative with respect to $k_x$  (or equivalently to $k_y$) at ${\bf k}=0$ and examining Eq.\,(\ref{op_0}) in the vicinity of ${\bf k}=0$ by a similar parabolic approximation for $\Delta_{\bf k}\simeq \Delta_0+ k^2 \Delta_0^{\prime\prime}/2$ where the first order terms in both expansions are absent due to the angular symmetry, we have a self consistency condition for $\Delta_0=\Delta_{\bf k}\vert_{\bf k=0}$ and $\Delta_{0}^{\prime\prime}=\Delta_{\bf k}^{\prime\prime} \vert_{\bf k=0}$ given by 
\begin{eqnarray}
\Delta_{0}&=&-E_0\{G_0+\frac{2}{D^2}G_0^{\prime \prime}\} ,\qquad E_0=\frac{e^2}{2\epsilon D}\nonumber \\
\Delta_{0}^{\prime \prime}&=&-\frac{\Delta_0}{\mu_x}\frac{\hbar^2}{m}
\label{laplace_2}
\end{eqnarray}
We notice that Eq.\,(\ref{op_0}) is easy to solve at the zero temperature since 
\begin{eqnarray}
\lim_{k\to 0}\,F(\epsilon_{k}^{(-)},E_{k})=\cases{1 \quad {\rm for} \quad \Delta_0^2+\mu_x^2 < \mu_-^2 \cr -1 \quad {\rm for} \quad \Delta_0^2+\mu_x^2 > \mu_-^2}\,.
\label{laplace_2_1}
\end{eqnarray}
The first case in Eq.\,(\ref{laplace_2_1}) is allowed when there is a high electron hole imbalance, indicated by a sufficiently large $\mu_-$. For such large $\mu_-$ there is no non-zero solution for $\Delta_{\bf k}$, and indeed this is verified by our earlier numerical calculations (Fig.(3) in \cite{2}). If $\mu_-$ is zero or weak, a nonzero solution is allowed by the lower case in Eq.\,(\ref{laplace_2_1}). Considering $\mu_-=0$ we find that,
\begin{eqnarray}
\Delta_0 &=& \sqrt{E_0^2-\mu_x^2}
\label{laplace_sol_3}
\end{eqnarray}
with $\Delta_{0}^{\prime \prime}$ as given by Eq.\,(\ref{laplace_2}). Eq.(\ref{laplace_sol_3}) is the first indication that the model can reproduce a sharp phase boundary. Indeed, at a critical layer separation $D=D_c$, $\Delta_0=0$ where $D_c=e^2/(2\epsilon \mu_x)$. A simple expansion of Eq.(\ref{laplace_sol_3}) near $D=D_c$ yields that, 
\begin{eqnarray}
\Delta_0\simeq \alpha \sqrt{1-\frac{D}{D_c}} ~\qquad {\rm with} \qquad \alpha=\sqrt{\frac{4}{3}}E_0~.   
\label{around_D_c}
\end{eqnarray}   
As a result of this parabolic expansion, we find that 
\begin{eqnarray}
E_{k} =\frac{E_{0}}{\mu_{x}} |\epsilon_{k}| 
\label{nrg}
\end{eqnarray} 
The exciton chemical potential $\mu_x$ is determined by the particle number conservation for $n_x=(n_e+n_h)/2$ as given by,  
\begin{eqnarray}
n_x=\frac{1}{A}\sum_{\bf k}(1-\frac{\epsilon_{\bf k}}{E_{\bf k}})
\label{number_1}
\end{eqnarray}
Solving this for $\mu_x$ using the parabolic approximation, 
\begin{eqnarray} 
\mu_x&=&-\frac{E_0}{2}+\sqrt{\Bigl(\frac{E_0}{2}\Bigr)^2+\frac{E_0 n_x}{\Gamma}} 
\label{chempot2} 
\end{eqnarray}  
where $\Gamma=m_x/(2\pi\hbar^2)$ is the two-dimensional density of states with $m_x$ being the exciton reduced mass. 
The condensation energy $\Delta \Omega$ given by Eq.\,(\ref{FE_1}) can be found similarly using Eq.'s(\ref{laplace_sol_3}) and (\ref{chempot2}). The result is, 
\begin{eqnarray}
\Delta \Omega=-\frac{\Gamma}{\,E_0} [\mu_x (\mu_x^2+\frac{3}{2}\Delta_0^2)-\mu_0^3]~,\qquad D \le D_c
\label{FE_2}
\end{eqnarray}
where $\mu_0=n_x/2\Gamma$ is the chemical potential $\mu_x$ evaluated at the phase boundary $\Delta_0=0$. Using Eq.(\ref{around_D_c}) and $\mu_x$ in Eq.(\ref{chempot2}), Eq.(\ref{FE_2}) can be represented at the phase boundary as
\begin{eqnarray}
\Delta \Omega=-3\Gamma \mu_0^2 (1-\frac{D}{D_c})~,\qquad D \simeq D_c
\label{Omega_at_D_c}
\end{eqnarray}
predicting a linear dependence of $\Delta \Omega$ with respect to $D$. A comparison between earlier numerical results in \cite{2} in the vicinity of the phase boundary and Eq's.(\ref{around_D_c}) and (\ref{Omega_at_D_c}) are shown in Fig.(\ref{check_Delta_0_Omega_0}). The accuracy of this simple parabolic approximation in capturing the main features of the  numerical calculations in \cite{2} is quite remarkable. 
Encouraged by this, now we proceed to one of the main result of our work, i.e. finding an expression for the EC-force. Using Eq.(\ref{Omega_at_D_c}) in Eq.(\ref{Casimir_force}) we find that  
\begin{eqnarray}
\frac{{\cal F}_{EC}}{A}\simeq -\frac{3}{4}\frac{n_x^2}{\Gamma D_c}~,\qquad D \simeq D_c
\label{cas_force_fin}
\end{eqnarray}

where $A$ is the sample area and $D_c$ is the critical layer separation. The latter can be found from Eq.\,(\ref{laplace_sol_3}) for $\Delta_0=0$ as $D_c=e^2\Gamma/(\epsilon n_x)$. Eq.(\ref{cas_force_fin}) indicates that, for a sample size of $A\simeq 10^3 \mu m^2$ and a typical concentration of $n_x\simeq 3\times 10^{11} cm^{-2}$, the EC-force is ${\cal F}_{EC} \simeq 10^{-9} N$ which is quite measurable within experimentally available precision. 

Here, a brief discussion about the realibility of the Hartree-Fock mean field approximation is also necessary. This technique which is successfully used in the theory of superconductivity as well as in BEC, requires in this case the self consistent handling of the condensate's dark and bright components simultaneously. The fundamental Pauli exchange scatterings between two dark (bright) excitons into to bright (dark) ones is a correction to Hartree-Fock self consistent scheme due to the fluctuations induced by the exciton-exciton interactions. The exchange of two electrons in the scattering of two dark states yields two bright states and visa versa. Due to the detailed balance, a marginal concentration of bright states is therefore expected to be present in the dominantly dark condensate. However, the existence of bright states, costs positive energy in the presence of a strong dark condensate. Pauli exchange mechanism can therefore be important at the phase boundary where the dark condensate is weakened. The result of this effect is to smear out the sharpness of the transition in Fig(\ref{check_Delta_0_Omega_0}) at $D=D_c$ and weaken the EC-force. Considering the linear dependence of the free energy in the inset of Fig.(\ref{check_Delta_0_Omega_0}), this marginal effect can be avoided by staying away from the phase boundary at $D_c=e^2\Gamma/(\epsilon n_x)$. This can be done easily by controlling $n_x$ via the power of the pump laser such that $D/D_c < 1$.       
  
\begin{figure}
 \vspace{-1.0cm}
\includegraphics[scale=0.31,angle=0]{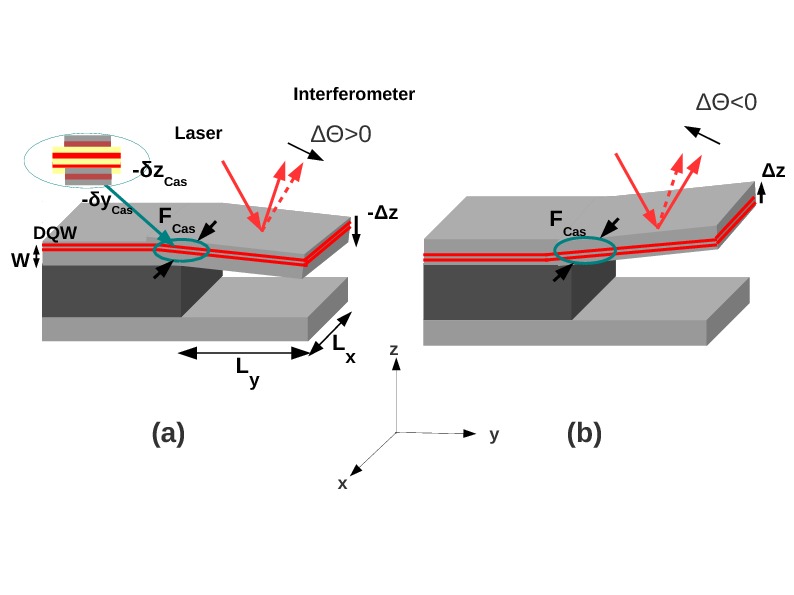}
 \vspace{-1.9cm}
\caption{(Color online) The proposed geometry for the static and dynamical measurements of the EC-force when the DQW is grown (a) above, and (b) below the neutral plane of the cantilever.}
\vspace{-0.5cm}
\label{casimir_setup}
\end{figure}

We are now at a point where we can discuss the experimental proposals to measure ${\cal F}_{EC}$. Due to the dielectric between the quantum wells, the direct measurement is much more challenging than measuring the electromagnetic Casimir Force between the two metallic plates separated by vacuum. To overcome this, we propose the experimental set-up in Fig.(\ref{casimir_setup}) based on the motion of a micro-mechanical resonator under a driving force. 
Here ${\cal F}_{EC}$ serves as a driving force for exciting static displacement as well as dynamical oscillations of the cantilever. Similar micro-mechanical oscillators with high quality factors have been successfully tested recently.\cite{Yamaguchi} 

For the measurement of the static displacements, the actuation of ${\cal F}_{EC}$ is to deform tetragonally the lattice around the EC within the DQW. If the DQW is grown asymmetrically with respect to the neutral plane of the cantilever, the emergence of ${\cal F}_{EC}$ yields a bending of the cantilever in accordance with Fig.(\ref{casimir_setup}). The deflection angle 
$\Delta \Theta$ can then be measured using a deflection-detection laser and an optical interferometer. Using classical theory of beam bending under stress\cite{Sokolnikoff}, we can estimate the angular deflection as $\Delta \Theta \simeq L_x L_y {\cal F}_{EC}/(12 E I)$ where $L_y \simeq 100 \mu m, L_x \simeq 10 \mu m$ as indicated in Fig.(\ref{casimir_setup}), E is the Young's modulus and $I\simeq L_y W^3/3$ is the second area moment of the curved structure with $W=0.3 \mu m$ being the thickness of the cantilever. Using $E\simeq 80 GPa$ in GaAs type materials, and for ${\cal F}_{EC} \simeq 10^{-9} N$, we find that $\Delta\theta \simeq 3 \times 10^{-4} rad$ which is possible to detect.\cite{AFM} 

A second and perhaps more realistic method is the measurement of the periodically driven dynamical oscillations of the cantilever. A typical cantilever oscillator\cite{Yamaguchi} can be driven with a power consumption $P=m_{eff} \Omega_0^3 \Delta z_{rms}^2/Q \simeq 2\times 10^{-15} J/s$ where $\Delta z_{rms}$ is the rms vibrational amplitude, $m_{eff}\simeq 10^{-10} kg$ is the effective mass of the cantilever, $\Omega_0/(2\pi)\simeq  20 kHz$ is the resonance frequency, and $Q\simeq 2.5\times 10^{5}$ is the quality factor of the resonator. In the Fig.(\ref{casimir_setup}) geometry at resonance, this leads to $\Delta z_{rms} \simeq 50 nm$ which is detectable. 

In order to be able to observe the predicted force using the scheme in Fig.(\ref{casimir_setup}), one has to take into account a delicate balance between various physical parameters such as the thermal relaxation time, thermomechanical noise, the pulse duration and the peak power of the pump laser in resonance with the band-width, the exciton lifetime and the photon pressure applied on the cantilever by the deflection-detection laser measuring the deflection angle. The created electrons and holes reach thermal equilibrium with the lattice within the thermal relaxation on the order of a few nanoseconds. On the other hand, the lattice thermalization time, which is important in consideration of the heating effects by the lasers, is much longer, in the order of a few microseconds. If the pulse period of the pump laser creating the electron-hole pairs (typically a few tens of kHz in mechanical resonance with the cantilever) is much shorter than the lattice thermalization time and if the peak power is within the mW range, the heating of the cantilever can be avoided. On the other hand, within a single period of the pump laser, the pairs are created and thermally relaxed within a few nanosceconds   and condensed in the ground state. During the idle part of the full period, the condensed excitons should disappear by recombination before the second pulse arrives. Therefore the pulse period should be much larger than the exciton lifetime. This is already a common practice in existing experiments considering the $\approx 10 \mu s$ lifetime of the excitons (in spatially indirect DQW geometry and under stabilizing static electric field) and a realistic $20-50 kHz$ pulse frequency of the pump laser. Once the condensate forms within each active pulse period, the appearance of the EC-force is almost instant in comparison with the pulse width. Therefore a pulse frequency of $20-50 kHz$ creates a periodic square-well shaped profile for the EC-force driven in the same frequency. The physical size of the caltilever is adopted after \cite{Yamaguchi} such that the natural resonance frequency of the cantilever is equal to the pulse frequency of the pump laser. As a result, the cantilever is brought in resonance with the periodic square-well modulations of the EC-force, very much like exciting a high quality harmonic spring periodically at its resonant frequency. The resulting resonant amplitude of the cantilever, considering the high quality factor $Q\simeq 2.5\times 10^{5}$ as in \cite{Yamaguchi}, turns out to be about $50 nm$. 

One should also be aware of another secondary effect, the photon force exerted by the deflection-detection laser in connection with the interferometric measurement of the cantilever's motion. If a mW range, $600 nm$ wavelength is used for the deflection-detection laser, a simple calculation shows that about $10^{-11} N$ constant photon force would be exerted on the cantilever at normal incidence, which is already a hundred times smaller than ${\cal F}_{EC}$. In order to avoid the heating of the cantilever by the deflection-detection laser, a highly reflecting thin coating can be made on the top part of the cantilever with a wide angle of incidence close to $\pi/2$. The wide angle also reduces the magnitude of the photon force by another $\approx 10^2$ times down to $10^{-12}-10^{-13} N$ range. The oscillation frequency as well as the amplitude of oscillations of the cantilever are however unaffected by this time independent and weak photon force (like harmonic spring oscillation frequency is unaffected by the constant gravity).         

Yet another important issue is minimizing the thermomechanical noise. This is an important factor usually for low resonant frequencies in the sub $kHz$ range which dies exponentially for higher resonant frequencies. This problem is handled in our case by a resonant frequency of the cantilever in the $20 kHz$ range.    

There are basically two main results of our work. The first result is that, an exciton condensate created in a DQW geometry, gives rise to a new type of force that is not known yet in other condensed matter systems. The EC-force, although it reminds the Casimir effect due to the electromagnetic vacuum fluctuations of the radiation field, here is the result of the Coulomb interaction. In the Casimir case, the driving mechanism is the dependence of the photon density of states on the geometrical scale imposed by the boundary conditions, whereas in the EC, it is the specific dependence of the Coulomb coupling on the geometry. As a result, the EC-force depends on the properties of the condensate as given by Eq.(\ref{cas_force_fin}). In particular the $1/D$ dependence in Eq.(\ref{cas_force_fin}) is contrasting with the $1/D^4$ dependence of the electromagnetic Casimir Force. A similar experimental set-up to our proposal in Fig(\ref{casimir_setup})  has already been tested in Ref.\cite{Yamaguchi} in a different context. Besides introducing this genuinely new force in physics as the first result, our second result in this work is not crucial to a lesser extend than the first one. This is motivated by the necessity to go beyond the standard photoluminescence experiments on EC in order to observe the true ground state properties dominated by the dark excitons. 

In this regard, the authors hope that, these results can motivate new generation experimental efforts complementary to the photoluminescence techniques on the EC. This work can also hopefully stimulate research in other quantum condensed systems where the conceptual mechanism of a force due to a quantum condensate can be investigated.  
\vspace{-0.3cm}  
\section{Acknowledgements}
\vspace{-0.3cm} 
The authors are grateful to Klaus-Juergen Friedland (Paul-Drude Institute, Germany) and Aykutlu D\^{a}na (UNAM, Bilkent University, Turkey) and Nai-Chang Yeh (Caltech, USA) for useful discussions concerning the experimental part of the work.   


\vspace{-0.3cm}   
\begin{thebibliography}{99}
\vspace{-0.3cm}
\bibitem{CF} H.B.G. Casimir, Proc. K. Ned. Akad. Wet. {\bf 51}, 793 (1948). 
\bibitem{early_Casimir} P. H. G. M van Blokland and J. T. G. Overbeek, J. Chem. Soc. Faraday Trans. I {\bf 74} , 2637 (1978); S. K. Lamoreaux  Phys. Rev. Lett. {\bf 78} , 5 (1997).
\bibitem{BCOR} G. Bressi, G. Carugno, R. Onofrio and G. Russo, Phys. Rev. Lett. {\bf 88}, 041804 (2002).
\bibitem{Casimir_theoretical_ways} V. M. Mostepanenko, N. N. Trunov, {\it The Casimir Effect and Its Applications} , Oxford Science Publications, (1997);  K. A. Milton, {\it The Casimir Effect} , World Scientific, (2001).
\bibitem{FG} M.E. Fisher and P.G. de Gennes, C.R. Acad. Sci. Paris, {\bf B 287}, 207 (1978).
\bibitem{HHGDB} C. Hertlein, L. Helden, A. Gambassi, S. Dietrich and C. Bechinger, Nature, {\bf 451}, 172 (2008); M. Fukuto, Y. F. Yano and P.S. Pershan, Phys. Rev. Lett. {\bf 94}, 135702 (2005); R. Garcia and M.H.W. Chan, Phys. Rev. Lett. {\bf 88}, 086101 (2002). 
\bibitem{BEC_CF_Th}  Shyamal Biswas, J K Bhattacharjee, Dwipesh Majumder, Kush Saha and Nabajit Chakravarty, J. Phys. {\bf B 43} (2010) 085305; P. A. Martin and V. A. Zagrebnov, Europhys. Lett. {\bf 73} 15 (2006).   
\bibitem{BEC_CF_Exp} J. M. Obrecht, R. J. Wild, M. Antezza, L. P. Pitaevskii, S. Stingari and E. A. Cornell,  Phys. Rev. Lett. {\bf 98}, 063201 (2007);  D. M. Harber, J. M. Obrecht, J. M. McGuirk and E. A. Cornell, Phys. Rev. A {\bf 72} 033610 (2005).
\bibitem{EC_general}  S.A. Moskalenko, Fiz. Tverd. Tela {\bf 4} , 276 (1962); J.M. Blatt, K.W. Böer, and W. Brandt, Phys. Rev. {\bf 126} , 1691 (1962).
\bibitem{EC_interactions_general} L.V. Butov, J. Phys.: Condens. Matter. {\bf 16}  R1577 (2004);
     L.V. Butov, J. Phys.: Condens. Matter. {\bf 19}  295202 (2007);
     D.W. Snoke, Adv. Condens. Matter. Phys. 2011 {\bf 1} (2011);
     D.W. Snoke, Science {\bf 298}   1368 (2002).
\bibitem{Combescot}  Monique Combescot, O. Betbeder-Matibet, Roland Combescot, Phys. Rev. Lett. {\bf 99}  176403 (2007).
\bibitem{1} M.A. Can and T. Hakioglu, Phys. Rev. Lett. {\bf 103} 086404 (2009); T. Hakioglu and M. Sahin, Phys. Rev. Lett. {\bf 98}, 166405 (2007). 
\bibitem{2} T. Hakio\u{g}lu and E. \"{O}zg\"{u}n, Sol. Stat. Comm. {\bf 151}, 1045 (2011). 
\bibitem{expl_2} Including realistically the difference  between the electron and the hole band masses yields a slightly asymmetric shape in the free energy as a function of $n_-$. Since we consider $n_-$ only at a single value in this work, i.e. $n_-=0$, the electron-hole mass difference is not crucial for the physics that follows.
\bibitem{Yamaguchi} H. Yamaguchi, H. Okamoto, S. Ishihara, and Y. Hirayama, App. Phys. Lett. {\bf 100}, 012106 (2012).
\bibitem{Sokolnikoff} I.S. Sokolnikoff, Mathematical Theory of Elasticity, Mc-Graw Hill, New York 1956.
\bibitem{AFM} The resolution in $\Delta \Theta$ is expected to be similar to the radian accuracy of the conventional AFMs of $\simeq 10^{-5} rad$. [see Dawn A. Bonnell et al., Rev. Mod. Phys. {\bf 84}, 1343 (2012).] 
\end{thebibliography}
\end{document}